\documentclass{article}
\usepackage{amsmath,amssymb,portland}
\def\cole{{\alpha}}
\def\kohl{{\beta}}
\def\davi{{\gamma}}
\def\tdeby{{\tau}}
\def\tcole{{\tau_\cole}}
\def\tkohl{{\tau_\kohl}}
\def\tdavi{{\tau_\davi}}
\def\tHN{{\tau_{H}}}
\def\u{u}
\def\i{\mbox{i}}

\def\LT#1{{\cal L}\left\{#1\right\}}

\def\HH#1#2{\left|\begin{array}{l}{#1}\\[1ex]{#2}\end{array}\right.}

\def\d{\mbox{\rm d}}

\def\f{f}

\def\o{\omega}

\DeclareMathOperator{\Rea}{Re}

\def\L1loc{L^1_{\rm loc}}

\def\Ga{\Gamma}

\def\Lc{{\cal L}}

\def\dst{\displaystyle}

\DeclareFontFamily{U}{msb}{}
\DeclareFontShape{U}{msb}{m}{n}{
<5><6><7><8><9> gen *msbm <10><10.95><12><14.4><17.28><20.74><24.88>msbm10}{}
\DeclareSymbolFont{AMSb}{U}{msb}{m}{n}
\DeclareMathSymbol{\bA}{\mathbin}{AMSb}{'101}
\DeclareMathSymbol{\bB}{\mathbin}{AMSb}{'102}
\DeclareMathSymbol{\bC}{\mathbin}{AMSb}{'103}
\DeclareMathSymbol{\bD}{\mathbin}{AMSb}{'104}
\DeclareMathSymbol{\bE}{\mathbin}{AMSb}{'105}
\DeclareMathSymbol{\bF}{\mathbin}{AMSb}{'106}
\DeclareMathSymbol{\bG}{\mathbin}{AMSb}{'107}
\DeclareMathSymbol{\bH}{\mathbin}{AMSb}{'110}
\DeclareMathSymbol{\bI}{\mathbin}{AMSb}{'111}
\DeclareMathSymbol{\bJ}{\mathbin}{AMSb}{'112}
\DeclareMathSymbol{\bK}{\mathbin}{AMSb}{'113}
\DeclareMathSymbol{\bL}{\mathbin}{AMSb}{'114}
\DeclareMathSymbol{\bM}{\mathbin}{AMSb}{'115}
\DeclareMathSymbol{\bN}{\mathbin}{AMSb}{'116}
\DeclareMathSymbol{\bO}{\mathbin}{AMSb}{'117}
\DeclareMathSymbol{\bP}{\mathbin}{AMSb}{'120}
\DeclareMathSymbol{\bQ}{\mathbin}{AMSb}{'121}
\DeclareMathSymbol{\bR}{\mathbin}{AMSb}{'122}
\DeclareMathSymbol{\bS}{\mathbin}{AMSb}{'123}
\DeclareMathSymbol{\bT}{\mathbin}{AMSb}{'124}
\DeclareMathSymbol{\bU}{\mathbin}{AMSb}{'125}
\DeclareMathSymbol{\bV}{\mathbin}{AMSb}{'126}
\DeclareMathSymbol{\bW}{\mathbin}{AMSb}{'127}
\DeclareMathSymbol{\bX}{\mathbin}{AMSb}{'130}
\DeclareMathSymbol{\bY}{\mathbin}{AMSb}{'121}
\DeclareMathSymbol{\bZ}{\mathbin}{AMSb}{'132}

\begin{document}
\setcounter{page}{0}
\begin{center}
{\bf\Large Analytical representations 
for relaxation functions of glasses}\\[24pt]
{\normalsize 
R. Hilfe$\mbox{\rm r}^{1,2}$\\[12pt]
$\mbox{ }^1$ICA-1, Universit{\"a}t Stuttgart,
Pfaffenwaldring 27, 70569 Stuttgart\\
$\mbox{ }^2$Institut f{\"u}r Physik,
Universit{\"a}t Mainz,
55099 Mainz, Germany}
 \end{center}
 \thispagestyle{empty}
 ~\\[2cm]
 \begin{abstract}
 Analytical representations in the time and frequency
 domains are derived for the most frequently used
 phenomenological fit functions for non-Debye 
 relaxation processes.
 In the time domain the relaxation functions corresponding to the
 complex frequency dependent Cole-Cole,
 Cole-Davidson and Havriliak-Negami 
 susceptibilities are also represented in 
 terms of $H$-functions.
 In the frequency domain the complex frequency dependent
 susceptibility function corresponding to the 
 time dependent stretched exponential
 relaxation function is given in terms of $H$-functions.
 The new representations are useful for fitting
 to experiment.
 \end{abstract}
 PACS: 77.22.Gm,61.20.Lc,71.55.Jv, 67.40.Pf, 78.30.Ly\\[3cm]
{\tt published in:\\Journal of Non-Crystalline Solids, 
vol. 305 (2002), page 122}
 \newpage
 Analytical representaions of relaxation functions
 in the time domain and susceptibilities in the
 frequency domain are important to fit experimental
 data in a broad variety of experiments on glasslike
 systems.
 Dielectric spectroscopy, viscoelastic modulus measurements,
 quasielastic light scattering, shear modulus and shear
 compliance as well as specific heat measurements
 all show strong deviations from the normalized
 exponential Debye relaxation function
 \begin{equation}
 \f(t)=\exp(- t/\tau)
 \label{expon} 
 \end{equation}
 where  $\tau$ is the relaxation time
 \cite{BNAP93}.
 All relaxation functions in this paper are normalized
 to $f(0)=1$.
 Relaxation in the frequency domain is described
 in terms of the normalized complex susceptibility 
 \begin{equation}
 \widehat{\chi}(\u)=\frac{\chi(\o)-\chi_\infty}{\chi_0-\chi_\infty}=
 1-\u\LT{\f(t)}(\u)
 \label{suslapl}
 \end{equation}
 where $\u=-\i\o$, $\o$ is the frequency, 
 $\chi(\o)$ is a dynamic susceptibility normalized
 by the corresponding isothermal susceptibility,
 $\chi_0=\lim_{\o\to 0}\Rea\chi(\o)$ is the static
 susceptibility,
 $\chi_\infty=\lim_{\o\to\infty}\Rea\chi(\o)$
 gives the ``instantaneous'' response,
 and $\LT{\f(t)}(\u)$ is the Laplace transform of the
 relaxation function $\f(t)$.
 For the exponential relaxation function this leads to
 \begin{equation}
 \widehat{\chi}(\o)=\frac{1}{1+i\o \tau},
 \label{debye}
 \end{equation}
 i.e. the well known Debye susceptibility.

 Most generalizations of equations \eqref{expon} and \eqref{debye}
 for glasses and other complex materials are obtained by the
 method of introducing a fractional ``stretching'' exponent.
 In the time domain this method leads to the ``stretched
 exponential'' or Kohlrausch relaxation function given as
 \begin{equation}
 \f(t)=\exp[-(t/\tkohl)^\kohl]
 \label{kohlrausch}
 \end{equation}
 with fractional exponent $\kohl$ \cite{WW70}.
 Of course all formulae obtained by the method of stretching
 exponents are constructed such that they reduce to the
 exponential Debye expression when
 the stretching exponent becomes unity.
 Extending the method of stretching exponents to
 the frequency domain one obtains
 the Cole-Cole susceptibility \cite{CC41}
 \begin{equation}
 \widehat{\chi}(\o)=\frac{1}{1+(i\o \tcole)^\cole},
 \label{colecole}
 \end{equation}
 the Davidson-Cole expression \cite{DC51}
 \begin{equation}
 \widehat{\chi}(\o)=\frac{1}{(1+i\o \tdavi)^\davi}
 \label{davidson}
 \end{equation}
 or the combined Havrialiak-Negami form \cite{HN66}
 \begin{equation}
 \widehat{\chi}(\u)=
 \frac{1}{(1+(\u \tHN)^\cole)^\davi}
 \label{havrneg}
 \end{equation}
 as empirical expressions for the experimentally
 observed broadened relaxation peaks.
 Most surprisingly, the analytical transformations
 between the time and frequency domain for general
 values of the parameters in these simple analytical
 expressions seem to be largely unknown \cite{AAC91}, and
 authors working 
 in the time domain usually employ the stretched exponential
 function while authors working in the frequency domain
 use the stretched susceptibilities.
 An exception are the results in \cite{GN93} where
 the real and imaginary part of the elastic modulus
 were obtained for Kohlrausch relaxation.
 Note however that there is a sign error in the
 real part in the results of \cite{GN93}.
 It is therefore the purpose of this short communication
 to rederive expressions for the Kohlrausch
 susceptibility in the frequency domain.
 Secondly the same methods are used to obtain for the first
 time the relaxation function corresponding to the Havriliak-Negami
 susceptibility (and hence also the Cole-Davidson and Cole-Cole
 susceptibilities) in the time domain.
 It is hoped that these expressions will be useful for
 facilitating the fitting of experimental data.

 The objective of this paper is achieved by employing 
 a method based on so called $H$-functions \cite{fox61}.
 The $H$-function of order $(m,n,p,q)\in\bN^4$ and with parameters
 $A_i\in\bR_+ (i=1,\ldots,p)$, $B_i\in\bR_+ (i=1,\ldots,q)$, 
 $a_i\in\bC (i=1,\ldots,p)$, and $b_i\in\bC (i=1,\ldots,q)$
 is defined for $z\in\bC,z\neq 0$ by a contour integral
 in the complex plane \cite{fox61,PBM90}
 \begin{equation}
 H^{m,n}_{p,q}\left(z
 \HH{(a_1,A_1),\ldots,(a_p,A_p)}{(b_1,B_1),\ldots,(b_q,B_q)}
 \right) =
 \frac{1}{2\pi i}\int_\Lc \eta(s)
 z^{-s}\;\d s
 \label{Hdef}
 \end{equation}
 where the integrand is
 \begin{equation}
 \eta(s)=
 \frac{\dst\prod_{i=1}^m\Ga(b_i+B_is)\prod_{i=1}^n\Ga(1-a_i-A_is)}
 {\dst\prod_{i=n+1}^p\Ga(a_i+A_is)\prod_{i=m+1}^q\Ga(1-b_i-B_is)} .
 \label{Hkrn}
 \end{equation}
 In (\ref{Hdef}) $z^{-s}=\exp\{-s\log|z|-i\arg z\}$ and $\arg z$
 is not necessarily the principal value.
 The integers $m,n,p,q$ must satisfy
 \begin{equation}
 0\leq m \leq q, \qquad 0\leq n \leq p ,
 \end{equation}
 and empty products are interpreted as being unity.
 For the conditions on the other parameters
 and the path of integration the reader is referred
 to the literature \cite{fox61} (see \cite[p.120ff]{hil98e}
 for a brief summary).
 The importance of these functions for glassy relaxation
 arises from the facts that (i) they contain most special
 functions of mathematical physics as special cases
 and (ii) their Laplace transform is again an $H$-function.
 Moreover they possess series expansions that are
 generalizations of hypergeometric series.

 Based on the convenient properties of $H$-functions
 it is possible to derive time and frequency domain
 expressions for all non-Debye relaxation functions
 and suceptibilities.
 The results of the calculations are summarized in the tables below.
 Table \ref{relaxationH} gives $H$-function representations
 for all relaxation functions in the time domain.
 Table \ref{relaxationS} gives the corresponding power series 
 for all relaxation functions for small and large times.
 Table \ref{susceptH} summarizes $H$-function representations
 for the susceptibilities in the frequency domain, and
 Table \ref{susceptS} gives their power series expansions.
 Note that those power series where the domain of
 validity is given by a limit are asymptotic series.
 In these tables the notation
 \begin{equation}
 \Ga(a,x)=\int_a^\infty y^{a-1}e^{-y}\d y
 \end{equation}
 denotes the complementary incomplete Gamma function,
 and the abbreviation
 \begin{equation}
 E_a(x) = \sum_{k=0}^\infty \frac{x^k}{\Ga(a k +1)}
 \end{equation}
 is the Mittag-Leffler function.
 In addition the short hand notation
 \begin{equation}
 H_a(x)=H^{11}_{11}\left(-x\HH{(1,1)}{(1,a)}\right)
 \end{equation}
 was introduced for writing the Kohlrausch susceptibility.

 In summary the present paper has given unified representations
 of nonexponential relaxation and non-Debye susceptibilities
 in terms of $H$-functions.
 These representations lead to computable expressions that
 can be used to investigate the relations between the
 Kohlrausch susceptibility and other fit functions \cite{hil01}.
 It is hoped that
 the $H$-function representations given here will help to
 facilitate the computational transformation 
 between the frequency and time domain in
 theoretical considerations and experiment.

\begin{table}
\caption{$H$-function representation for the 
normalized relaxation functions ($\f(0)=1$) with relaxation
time $\tau$.}
\begin{tabular}{|l|c|c|}
&$f(t)$& $H$-function  \\[6pt] \hline
Debye & $\exp(-t/\tdeby)$ &
$\dst H^{10}_{01}\left(\frac{t}{\tdeby}\HH{-}{(0,1)}\right)$ 
\\[12pt] \hline
Kohlrausch & $\exp(-(t/\tkohl)^\kohl)$ &
$\dst H^{10}_{01}\left(\left(\frac{t}{\tkohl}\right)^\kohl
\HH{-}{(0,1)}\right)$ 
\\[12pt] \hline
Cole-Cole & $E_\cole(-(t/\tcole)^\cole)$ &
$\dst H^{11}_{12}\left(\left(\frac{t}{\tcole}\right)^\cole
\HH{(0,1)}{(0,1)(0,\cole)}\right)$ 
\\[12pt] \hline
Cole-Davidson& $\dst\frac{\Ga(\davi,t/\tdavi)}{\Ga(\davi)}$ &
$1-\dst\frac{1}{\Ga(\davi)}
H^{11}_{12}\left(\frac{t}{\tdavi}
\HH{(1,1)}{(\davi,1)(0,1)}\right)$ 
\\[12pt] \hline
Havriliak-Negami &  &
$1-\dst\frac{1}{\Ga(\davi)}
H^{11}_{12}\left(\left[\frac{t}{\tHN}\right]^\cole
\HH{(1,1)}{(\davi,1)(0,\cole)}\right)$ 
\end{tabular}
\label{relaxationH}
\end{table}

\newpage

\begin{table}
\caption{Series expansions for normalized relaxation 
functions ($\f(0)=1$) with relaxation time $\tau$.
Series are asymptotic whenever its range of validity is given as a limit.}
\begin{tabular}{|l|c|cr|}
&$f(t)$& series&\\[6pt] \hline
Debye & $\exp(-t/\tdeby)$ &
$\dst\sum_{k=0}^\infty\frac{(-1)^k}{\Ga(k+1)}
\left(\frac{t}{\tdeby}\right)^k$
&$\dst\frac{t}{\tdeby}<\infty$\\[12pt]
&&$\exp(-t/\tdeby)$
&$\dst\frac{t}{\tdeby}\to\infty$\\[12pt] \hline
Kohl- & $\exp(-(t/\tkohl)^\kohl)$ &
$\dst\sum_{k=0}^\infty\frac{(-1)^k}{\Ga(k+1)}
\left(\frac{t}{\tkohl}\right)^{\kohl k}$
&$\dst\frac{t}{\tkohl}<\infty$\\[12pt]
rausch&&$\exp(-(t/\tkohl)^\kohl)$
&$\dst\frac{t}{\tkohl}\to\infty$\\[12pt] \hline
Cole- & $E_\cole(-(t/\tcole)^\cole)$ &
$\dst\sum_{k=0}^\infty\frac{(-1)^k}{\Ga(\cole k+1)}
\left(\frac{t}{\tdeby}\right)^{\cole k}$
&$\dst\frac{t}{\tcole}<\infty$\\[12pt]
Cole&&$\dst\sum_{k=1}^\infty
\frac{(-1)^{k+1}}{\Ga(1-\cole k)}
\left(\frac{t}{\tcole}\right)^{-\cole k}$
&$\dst\frac{t}{\tcole}\to\infty$\\[12pt] \hline
Cole-& $\dst\frac{\Ga(\davi,t/\tdavi)}{\Ga(\davi)}$ &
$\dst 1-\frac{1}{\Ga(\davi)}\sum_{k=0}^\infty\frac{(-1)^k}{(k+\davi)\Ga(k+1)}
\left(\frac{t}{\tdavi}\right)^{k+\davi}$
&$\dst\frac{t}{\tdavi}<\infty$\\[12pt]
Davidson&&
$\dst\frac{\exp(-t/\tdavi)}{\Ga(\davi)}
\left(\frac{t}{\tdavi}\right)^{\davi-1}\left[
1+\sum_{k=0}^\infty\prod_{j=1}^k(\davi-j)
\left[\frac{t}{\tdavi}\right]^{-k}
\right]$
&$\dst\frac{t}{\tdavi}\to\infty$\\[12pt] \hline
Havriliak- &  &
$\dst1-\frac{1}{\Ga(\davi)}\sum_{k=0}^\infty
\frac{(-1)^k\Ga(k+\davi)}{\Ga(\cole k+\cole\davi+1)\Ga(k+1)}
\left[\frac{t}{\tHN}\right]^{\cole(k+\davi)}$&
$\dst\frac{t}{\tHN}<\infty$\\[12pt]
Negami&&$\dst\frac{1}{\Ga(\davi)}\sum_{k=1}^\infty
\frac{(-1)^{k+1}\Ga(k+\davi)}{\Ga(1-\cole k)\Ga(k+1)}
\left(\frac{t}{\tHN}\right)^{-\cole k}$&$\cole\neq 1$,
$\dst\frac{t}{\tHN}\to \infty$
\end{tabular}
\label{relaxationS}
\end{table}

\newpage

\begin{table}
\caption{H-function representations for the normalized 
frequency dependent complex susceptibilities ($\u=-\i\o$).}
\begin{tabular}{|l|c|c|}
& $\widehat{\chi}(\u)$ & $H$-function \\[6pt] \hline
Debye & $\dst \frac{1}{1+\u\tdeby}$ &
$\dst H^{11}_{11}\left(\u\tdeby\HH{(0,1)}{(0,1)}\right)$ \\[12pt] \hline
Kohlrausch & $1-H_\kohl(-(\u\tkohl)^\kohl)$ &
$\dst 1-H^{11}_{11}\left((\u\tkohl)^\kohl
\HH{(1,1)}{(1,\kohl)}\right)$ 
\\[12pt] \hline
Cole-Cole & $\dst \frac{1}{1+(\u\tcole)^\cole}$ &
$\dst H^{11}_{11}\left((\u\tcole)^\cole\HH{(0,1)}{(0,1)}\right)$ 
\\[12pt] \hline
Cole-Davidson & $\dst \frac{1}{(1+\u\tdavi)^\davi}$ &
$\dst \frac{1}{\Ga(\davi)}H^{11}_{11}\left(\u\tdavi
\HH{(1-\davi,1)}{(0,1)}\right)$ 
\\[12pt] \hline
Havriliak-Negami & $\dst \frac{1}{(1+(\u\tHN)^\cole)^\davi}$ &
$\dst \frac{1}{\Ga(\davi)}H^{11}_{11}\left((\u\tHN)^\cole
\HH{(1-\davi,1)}{(0,1)}\right)$ 
\end{tabular}
\label{susceptH}
\end{table}

\newpage

\begin{table}
\caption{Series representations for the normalized 
frequency dependent complex susceptibilities ($\u=-\i\o$).
Series are asymptotic whenever its range of validity is given as a limit.}
\begin{tabular}{|l|c|cr|}
& $\widehat{\chi}(\u)$  & series&\\[6pt] \hline
Debye & $\dst \frac{1}{1+\u\tdeby}$ &
$\dst\sum_{k=0}^\infty  (-1)^k(\u\tdeby)^k$ & $|\u\tdeby|<1$ \\[12pt]
&&$-\dst\sum_{k=0}^\infty  (-1)^k(\u\tdeby)^{-k-1}$ & $|\u\tdeby|>1$ 
\\[12pt] \hline
Kohlrausch & $1-H_\kohl(-(\u\tkohl)^\kohl)$ &
$\dst 1-\sum_{k=0}^\infty\frac{(-1)^k\Ga((k+1)/\kohl)}{\kohl\Ga(k+1)}
(\u\tkohl)^{k+1}$&$|\u\tkohl|\to 0$\\[12pt]
&&$\dst 1-\sum_{k=0}^\infty\frac{(-1)^k\Ga(\kohl k+1)}{\Ga(k+1)}
(\u\tkohl)^{-\kohl k}$&$|\u\tkohl|> 0$\\[12pt] \hline
Cole-Cole & $\dst \frac{1}{1+(\u\tcole)^\cole}$ &
$\dst\sum_{k=0}^\infty (-1)^k (\u\tcole)^{\cole k}$&
$|\u\tcole|<1$\\[12pt]
&&$-\dst\sum_{k=0}^\infty (-1)^k (\u\tcole)^{-\cole(k+1)}$&
$|\u\tcole|>1$
\\[12pt] \hline
Cole-Davidson & $\dst \frac{1}{(1+\u\tdavi)^\davi}$ &
$\dst\sum_{k=0}^\infty\frac{(-1)^k\Ga(k+\davi)}{\Ga(\davi)\Ga(k+1)}
(\u\tdavi)^k$&$|\u\tdavi|<1$\\[12pt]
&&$-\dst\sum_{k=0}^\infty\frac{(-1)^k\Ga(k+\davi)}{\Ga(\davi)\Ga(k+1)}
(\u\tdavi)^{-(k+\davi)}$&$|\u\tdavi|>1$
\\[12pt] \hline
Havriliak-Negami & $\dst \frac{1}{(1+(\u\tHN)^\cole)^\davi}$ &
$\dst\sum_{k=0}^\infty\frac{(-1)^k\Ga(k+\davi)}{\Ga(\davi)\Ga(k+1)}
(\u\tHN)^{\cole k}$&$|\u\tHN|<1$\\[12pt]
&&
$-\dst\sum_{k=0}^\infty\frac{(-1)^k\Ga(k+\davi)}{\Ga(\davi)\Ga(k+1)}
(\u\tHN)^{-\cole(k+\davi)}$&$|\u\tHN|>1$
\end{tabular}
\label{susceptS}
\end{table}

\end{document}